\RequirePackage{fix-cm}
\documentclass[smallextended]{svjour3}

\smartqed

\usepackage{url}
\usepackage{nicefrac}
\usepackage{gensymb}
\usepackage{enumerate}
\usepackage{makecell}
\usepackage[utf8]{inputenc}
\usepackage{graphicx}
\usepackage{caption}
\usepackage{amssymb}
\usepackage{amsmath}
\usepackage{subfig}

\journalname{Experimental Astronomy}

\begin{document}

\title{Positioning optimization of a low cast portable star tracker up to 200 meters accuracy}
\author{Meysam~Izadmehr, Mehdi~Khakian~Ghom }
\institute{Meysam~Izadmehr, Mehdi~Khakian~Ghom \at Department of Energy Engineering and Physics, Amirkabir University of Technology, 15875-4413, Tehran, Iran}

\maketitle

\begin{abstract}

Using a portable star tracking system, we are able to obtain the geographical position of the observer, longitude, latitude and North direction. In this work, a new optimization method has been applied to improve accuracy in calculation of the observer position better than 6 arcsec (200 meters). In this method, we used 80 to 100 stars in each taken picture from the sky to apply optimization method. To obtain the accuracy it has been recorded observations for frequent 50 nights. In each night it has been taken 100 images.

\end{abstract}


\section{Introduction}
Today finding geographical position is a normal subject. Every smart mobile phone has the accessibility and finds our geographical longitude, latitude and height from sea level up to few ten meters accuracies \cite{hulbert2001accuracy,wing2005consumer}. But as we know and we have had the experience in our real life; technology makes our life much easier than before. But the problem is when we become familiar with a technology and depend on it. In this situation when we cannot have access to the technology we are not able to do anything; only we have to wait for solving the problem and continue our normal life. But always we need to have some alternatives which are more independent of the technology and usually the most important reference is nature near around us \cite{pappalardi2001alternatives}. A part of the nature is the sky, where we don't pay attention to it at all, and it covers half of our observable solid angle.

Ancient times navigators were using stars as a reference to find their ways, right directions \& positions. By development of the technology, the measurements have become more accurate \cite{secroun2001high}. Izadmehr et.al. \cite{IzadmehrArXiv180600607I} tried to improve the measurements, this report tries to optimize the errors by using as the most numbers of stars as possible in its taken pictures. Use a very light system to be a portable and low price instrument to obtain our position on the Earth coordinate system (CS) as accessible as it could be. Meanwhile, the direction of the north at the position of the observer is accessible by this method with an accuracy of 6 arcsec too. Benefits of the instrument are as follow: 
 \begin{enumerate}
   \item Providing angle with true north not magnetic north. 
    \item It's low cost. The electronic board is about 200\$. We need a camera with angular resolution better than 0.0025 degrees, which most of the cameras today have the accessibility.
    \item It is portable, the board and the camera and a laptop approximately about 3 to 4 kg.
    \item Need to a narrow field of view of less than 12 degrees. 
\end{enumerate}

The details of the instrument and its capabilities have been presented in Izadmehr et.al. \cite{IzadmehrArXiv180600607I}
. It is optimized the accuracy in this report by a new method which is based on the least square method by using about 80 to 100 stars in each taken picture from the sky. 

In chapter 2 it is presented the procedure for finding the position, its details, and sequential steps. Chapter 3 presents the calculation method of paper which is quite a new method to improve the accuracy of the results via least square method by using as the most number of stars as possible in each taken picture. In chapter 4 it is presented the result of 50 nights observations. Each night, it is taken 100 images from a part of the sky with the exposure time of 0.5 seconds. But 50 taken pictures in different 50 nights have been taken from different parts of the sky. By using the optimization method the accuracy of the positioning improved up to 50 times better. Chapter 5 concludes the results and the futures plans of this work.

\section{Positioning procedure}
The main goal of this section is calculation of an observed longitude and latitude $(\lambda, \varphi)$ using star pattern of the night sky. $\lambda$, $\varphi$ and the north direction are outputs of the following procedure:

\begin{equation}\label{eq:1}
A_2 A_3 \textbf{W} = A_4 A_1 \textbf{V}
\end{equation}

\begin{figure}
    \centering
    \subfloat[Configuration of the used coordinate systems. ]{%
    	\includegraphics[width=0.75\textwidth]{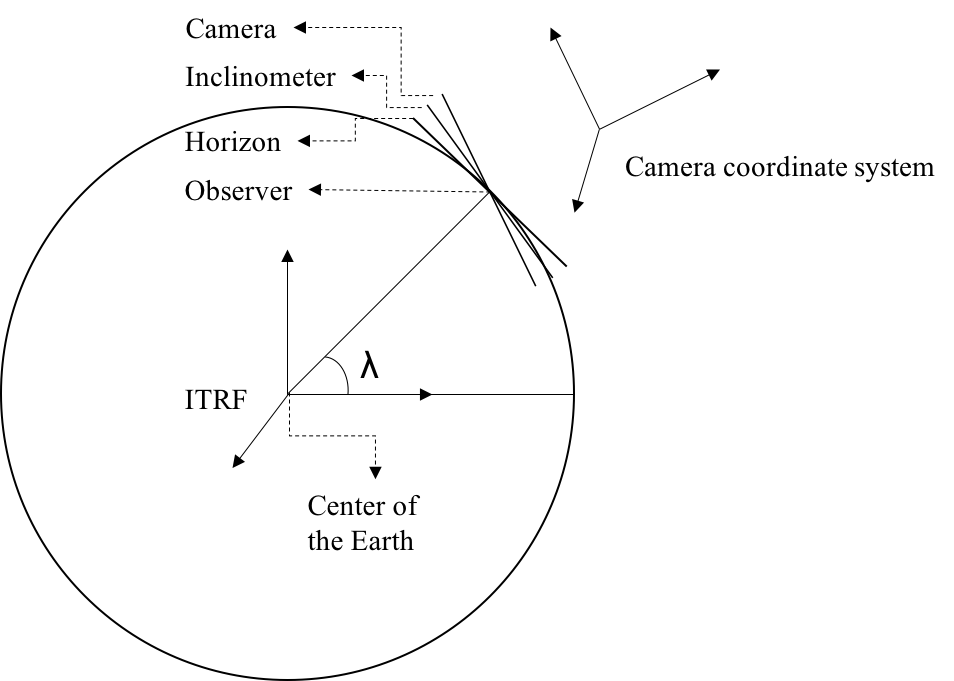} %
	}

    	\includegraphics[width=0.75\textwidth]{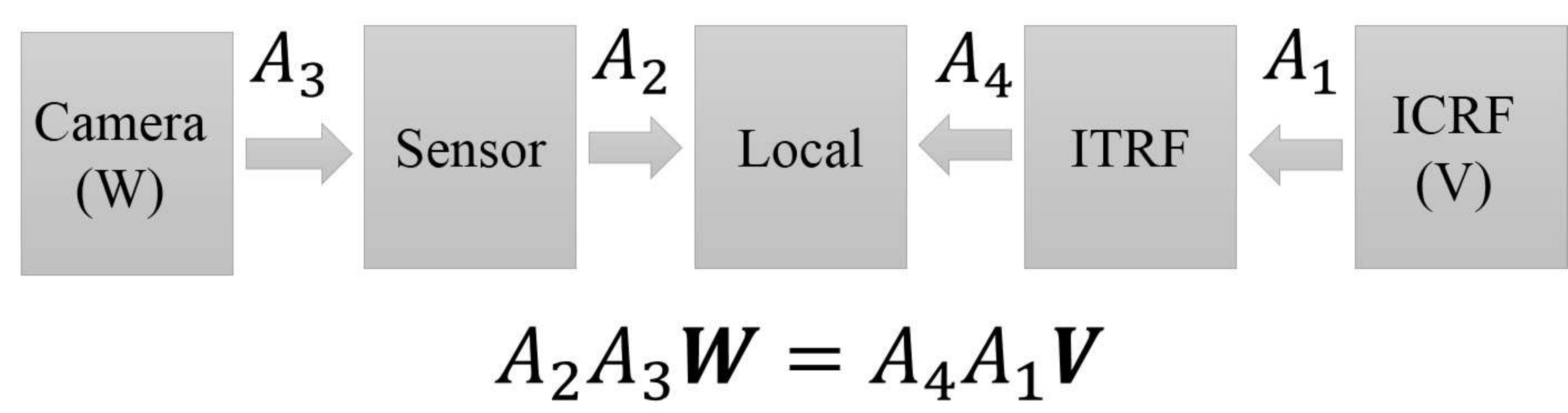}%
	
    \caption{
     \textbf{W} and \textbf{V} are unit vectors in camera CS and ICRF, respectively.\\\hspace{\textwidth}
$A_1$ is the rotational matrix which converts star vectors in ICRF into ITRF.\\\hspace{\textwidth}
 $A_2$ converts sensor CS into the local CS.\\\hspace{\textwidth}
$A_3$ converts camera CS into the sensor CS.\\\hspace{\textwidth}
 $A_4$ converts ITRF into the observer's local CS.}
    
     \label{fig:Schematic_design_of_coordinates_conversions}%
\end{figure}

Where $W$ and $V$ are unit 3 dimensional vectors in camera and reference CS respectively. $A_i$ are $3\times3$ rotational matrices for the projective of the two vectors from one system to another one (Figure~\ref{fig:Schematic_design_of_coordinates_conversions}). 

Matrix $A_1$ rotates ICRF\footnote{International Celestial Reference Frame} to ITRF\footnote{International Terrestrial Reference Frame}.
Matrix $A_4$ rotates ITRF to the observers CSs. (Figure~\ref{fig:Schematic_design_of_coordinates_conversions})

From left hand side of Eq.~\ref{eq:1}, it should project the camera taken picture to the observer's CS and result compares with the previous results. 

Matrix $A_3$ projects the taken picture from the night sky to the essential plane on the inclinometer CS. The optimization of the paper which improves the result of our positioning is in the matrix $A_3$. The uncertainty is due to the installation of the camera and inclinometer, which is dominant in the configuration system. Therefore we tried to decrease it by a least square method.

Matrix $A_2$ projects the gravity CS essential plane to the observer's CS. Inclinometer plane is projected to the horizontal plane of the observer by the output two angels of the inclinometer with the accuracy of 0.0025 degrees.

The main plane of the observer's local CS is on its horizontal plane, but the mismatching is the angle between their axis. the x-axis of the horizontal plane is to the north, and the angle between the x-axis of the observed CS and the north direction is the north angle which is one of the outputs of the system for the observer.

\section{Calculating rotational matrix between camera and inclinometer Coordinate System ( matrix $A_3$)}
Matrix $A_3$ could be minimized by a calibration procedure. The procedure is recursive. At first, matrix $A_4$ is calculated for latitude and longitude of the observer and then multiplied by $V$ vector to obtain the vector in the local CS. Angles of the rotation matrix between camera CS and local CS are calculated first. Then, using these angles, the camera CS is rotated to be projected on the horizon plane. The inclinometer is aligned to the plumb line using inclinometer outputs. This procedure is not the most accurate one. The best solution is to calculate the $A_3$ in a known location and use it as a constant. It means, star vectors $W$, $V$ are used for the known observer latitude and longitude. First, $A_4$ decomposed to its rotational matrices:

\begin{equation}\label{eq:A4}
A_4= R_3(c)R_2(\frac{\pi}{2}-\lambda)R_3(\beta)
\end{equation}

Where $\beta$ and $\lambda$ are longitude and latitude of the observer, respectively.
Using $A_4$, equation Eq.~\ref{eq:1} converts as follow:

\begin{equation}\label{eq:2}
R_3(-c)A_2 A_3 \textbf{W} = R_2(\frac{\pi}{2}-\lambda)R_3(\beta) A_1 \textbf{V}
\end{equation}

Latitude and longitude are available for the certain points, but $c$ depends on the camera direction. In the right side of Eq.~\ref{eq:2}, $A_1$, $V$, $\beta$ and $\lambda$ are quite available, therefore it becomes:
	
\begin{equation}\label{eq:3}
A' \textbf{W} =  \textbf{V$_{Alt-AZ}$}
\end{equation}

Where \textbf{V$_{Alt-AZ}$} is quite known vector in Horizontal CS and $A'$ is $R_3(-c)A_2 A_3$.  \textbf{V$_{Alt-AZ}$} and \textbf{W} are available star vectors for each star on the taken picture from the night sky. Therefore,  $A'$ could be optimized by the least square method, which has been described in section \ref{calculation_rotational_matrix_from_star_vectors}.

\subsection{calculation of the optimized rotational matrix from the star vectors} \label{calculation_rotational_matrix_from_star_vectors}

 The algorithm for calculating this rotational matrix divides into two categories. The first one uses a minimal set of data and then solves three possibilities for the nonlinear equations to obtain the altitude \cite{wang2014}. This category is generally called deterministic category,
a name which has been popularized by Wertz \cite{wertz2012}. The most well known deterministic algorithm in current use is the TRIAD algorithm \cite{shuster2004}. 

The rest of the algorithms, generally called optimal, which determines the altitude by minimizing
an appropriate cost function \cite{markley1993attitude} . In spite of the deterministic method which is using two vectors, the new method uses all extracted stars from the sky picture. 
The equation $A\textbf{W} = \textbf{V}$ can be solved by minimization of the non-negative equation \cite{wahba1965least}:

\begin{equation}\label{eq:4}
L(A) = \sum_{i=1}^{n} |\textbf{W}'_i - A\textbf{V}'_i|^2
\end{equation}

 Where $n$ is the number of the identified stars in the picture. Usually for a normal camera with aperture120 mm, focal length 600 mm and exposure time of 0.5 seconds in a normal sky with magnitude limitation up to 9.5. The number of the identified stars in the picture, $n$, is usually about 80 to100.
 \cite{IzadmehrArXiv180600607I}
Eq.~\ref{eq:4} can be broken down to:

\begin{equation}\label{eq:5}
 \begin{aligned}
L_x(A) = \sum_{i=1}^{n} |{W_x}'_i - (A_{0,0}{V'_x}_i+A_{0,1}{V'_y}_i+A_{0,2}{V'_z}_i)|^2\\
L_y(A) = \sum_{i=1}^{n} |{W_y}'_i - (A_{1,0}{V'_x}_i+A_{1,1}{V'_y}_i+A_{2,2}{V'_z}_i)|^2\\
L_z(A) = \sum_{i=1}^{n} |{W_z}'_i - (A_{2,0}{V'_x}_i+A_{2,1}{V'_y}_i+A_{2,2}{V'_z}_i)|^2
\end{aligned}
\end{equation}
These three equations are solved independently and each of them provides a row  of matrix$A$.  ALGIB \cite{bochkanov2011alglib} has been used for solving each independent equation of Eq.~\ref{eq:5}. ALGIB reduce the matrix to bidiagonal form and then diagonalize it using QR algorithm. This simple method is quite efficient, but it can speed up an algorithm significantly \cite{goodall199313}.

Matrix $A_3$ calculated by Eq.~\ref{eq:6}:

\begin{equation}\label{eq:6}
A' = R_3(-c)A_2A_3
\end{equation}

In Eq.~\ref{eq:6}, $A_2= R_2(-b)R_1(-a)$. $a$ and $b$ are rotational angels from sensor outputs. Therefore, $R_3(-c)A_2$ is equal to:

\begin{equation}\label{eq:7_1}
R_3(-c)A_2=
\left[ 
\begin{array}{ccc}
C_cC_b & C_cS_bS_a+S_cC_a & C_cS_bC_a-S_cS_a \\
-S_cC_b & -S_cS_bS_a+C_cC_a & -S_cS_b C_a-C_cS_a \\
-S_b & C_bS_a & C_bC_a
 \end{array} 
 \right]
\end{equation}
Where $C_i$s and $S_i$s are standing for $\cos(i)$ and $\sin(i)$ respectively. 
In Eq.~\ref{eq:6}, elements of the $A_3$ and $c$ angle are unknown. Because $c$ is unknown, all elements of the matrix product $R_3(-c)A_2A_3$ couldn't be used. The third row of the $R_3(-c)A_2$ is independent of the angle $c$ (Eq.~\ref{eq:7_1}). Three equations are extracted from Eq.~\ref{eq:6}:

\begin{equation}\label{eq:7}
 \begin{aligned}
 -S_bA_{0,0}+C_bS_aA_{1,0}+C_bC_aA_{2,0} = A'_{2,0}\\
  -S_bA_{0,1}+C_bS_aA_{1,1}+C_bC_aA_{2,1} = A'_{2,1}\\
   -S_bA_{0,2}+C_bS_aA_{1,2}+C_bC_aA_{2,2} = A'_{2,0}
\end{aligned}
\end{equation}
Eq.~\ref{eq:7}, shows three independent equations, with 9 unknown elements. Eq.~\ref{eq:7} can be converted to a matrix equation:

\begin{equation}\label{eq:8}
\left[ 
\begin{array}{ccc}
A_{0,0} & A_{1,0} & A_{2,0} \\
A_{0,1} & A_{1,1} & A_{2,1} \\
A_{0,2} & A_{1,2} & A_{2,2} 
 \end{array} 
 \right]
\times
 \left[ 
\begin{array}{c}
 -S_b\\
C_bS_a\\
C_bC_a
 \end{array} 
 \right]
 =
  \left[ 
\begin{array}{c}
A'_{2,0}\\
A'_{2,1}\\
A'_{2,0}
 \end{array} 
 \right]
\end{equation}
This equation could be written as: 

\begin{equation}\label{eq:9}
A'' \textbf{V}' =  \textbf{W}'
\end{equation}

Where $\textbf{V}' $ and $ \textbf{W}'$ obtain by the image data, stars catalog, latitude, longitude and inclinometer outputs. For each picture two vectors $\textbf{V}' $ and $ \textbf{W}'$ are created. Therefore, for a set of  $\textbf{V}' $ and $ \textbf{W}'$ vectors, from different images, matrix elements are obtained by the optimization method as described previously. 

\section {Results}
Results of positioning error for two situations have been used. One without calculating $A_3$ and just minimizing the $A_3$ by instrument, the other one is with the application of the least square method, calculating it with the presented method. Therefore latitude and longitude for 50 different locations and nights are used for both conditions. Each night 100 images have been taken in the same direction for the investigations. 
Average and standard deviation of this part of the sky for each night has been calculated. Since images from a part of the sky for each night has been taken in the same direction, standard deviation shows error populated by the calculation and image processing procedures. The average error indicated error populated by the inclinometer.

\begin{figure} 
\centerline{\includegraphics[width=0.7\textwidth]{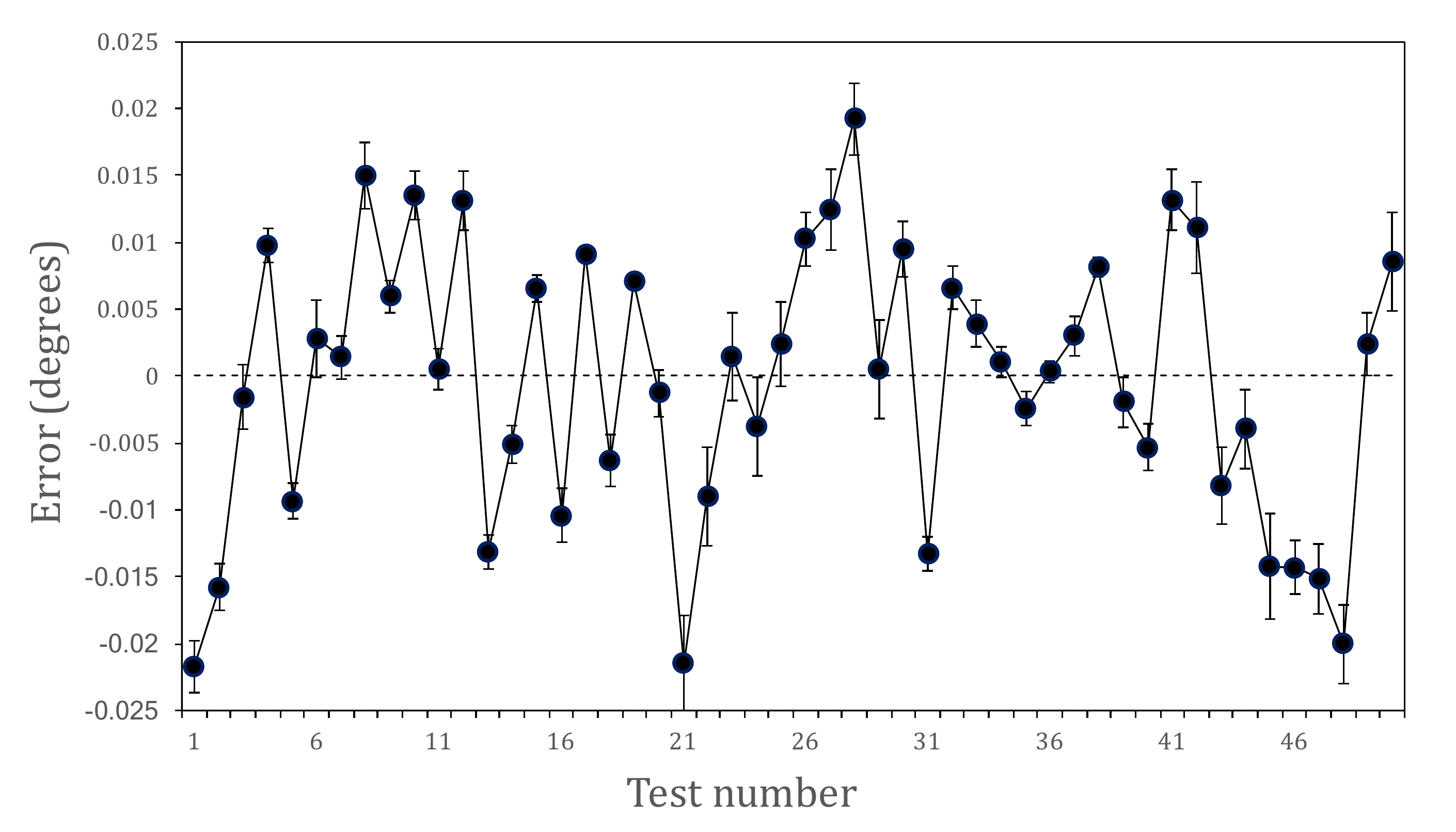}}
 \caption{Latitude error in 50 nights as well as different locations and camera directions. Average absolute deviation of latitude is  $0.5028$ arcminutes. Each point is obtained by at least the average of 100 images.
 \label{fig:old_latitude_error}}
\end{figure}

\begin{figure} 
\centerline{\includegraphics[width=0.7\textwidth]{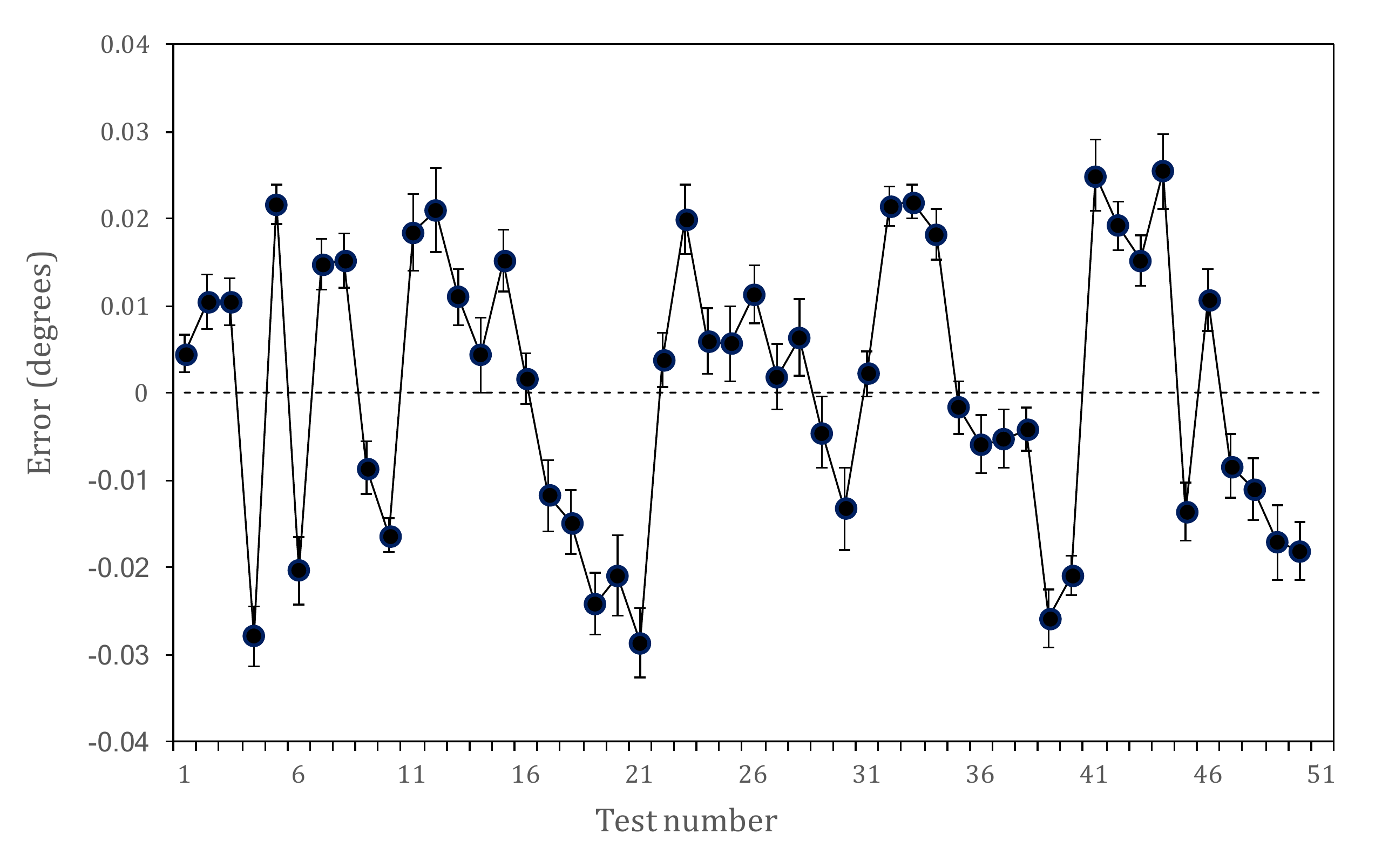}}
 \caption{Longitude error in 50 nights as well as different locations and camera directions. Average absolute deviation of longitude is $0.816$ arcminutes. Each point is obtained by at least the average of 100 images.
 \label{fig:old_longitude_error}}
\end{figure}

\begin{figure} 
\centerline{\includegraphics[width=0.7\textwidth]{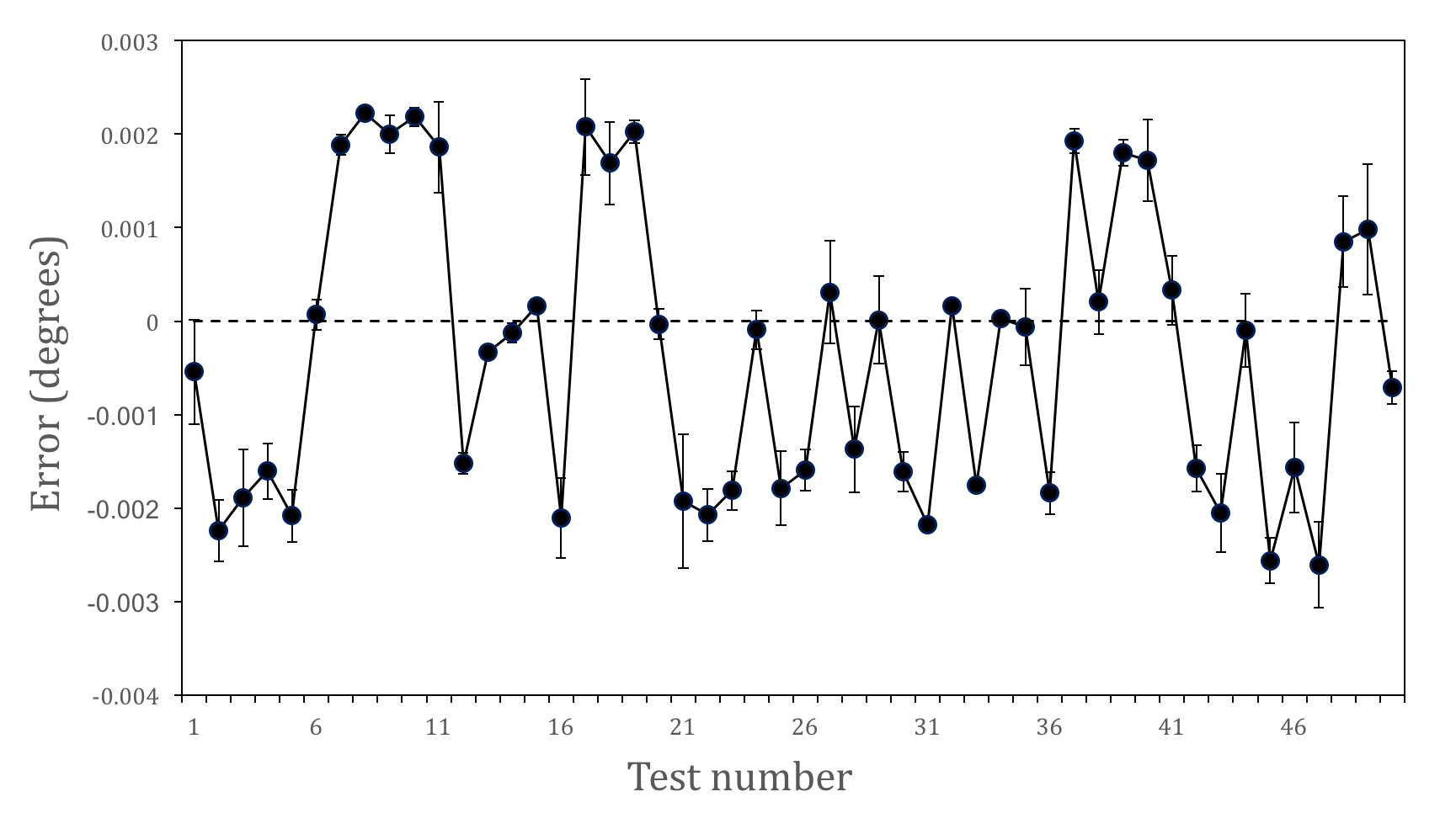}}
  \caption{Latitude error in 50 nights as well as different locations and camera directions. Average absolute deviation of latitude is  $4.77$ arcseconds. Each point is obtained by at least the average of 100 images.
 \label{fig:LatitudeResults}}
\end{figure}

\begin{figure} 
\centerline{\includegraphics[width=0.7\textwidth]{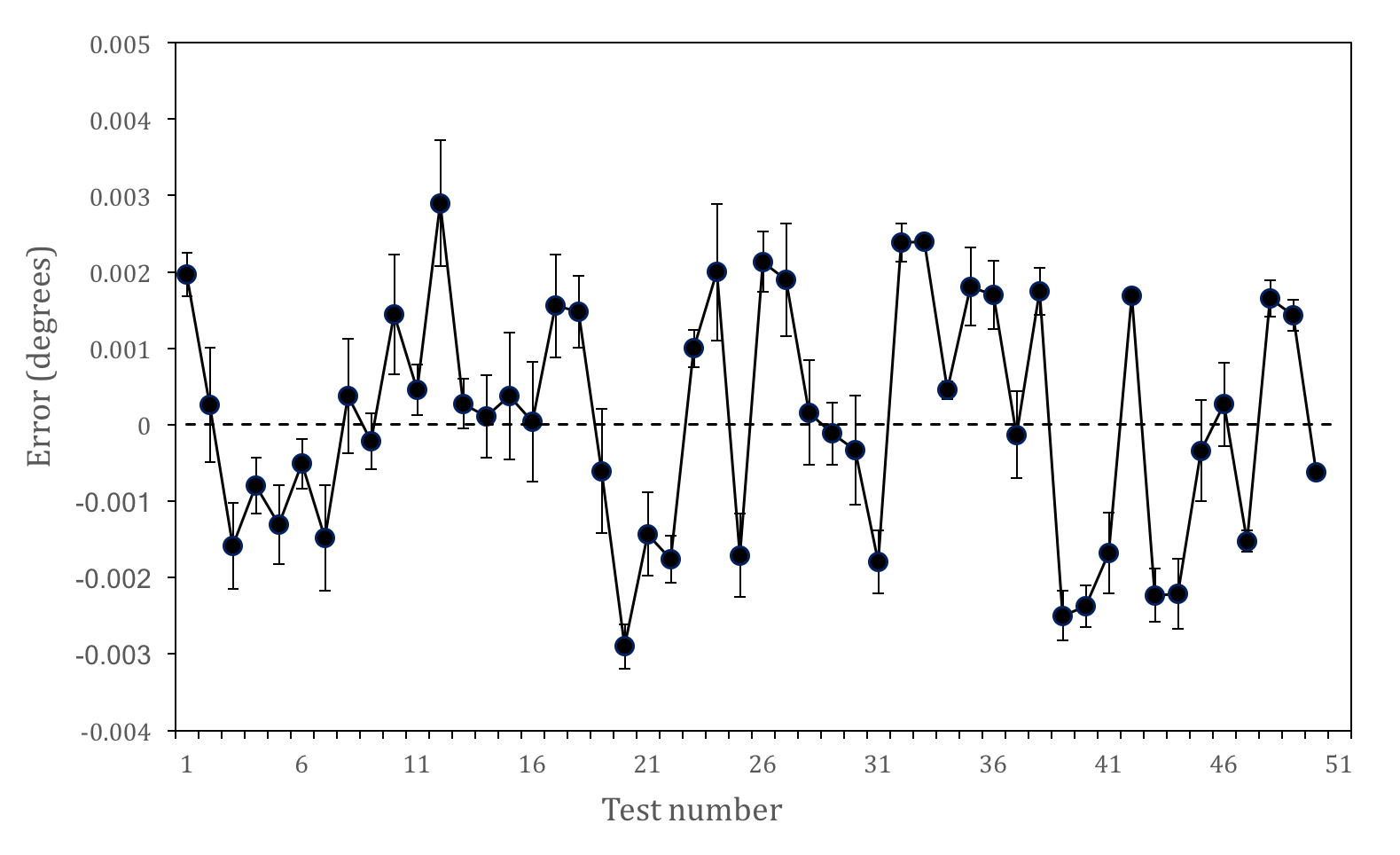}}
  \caption{Latitude error in 50 nights as well as different locations and camera directions. Average absolute deviation of latitude is  $5.53$ arcseconds. Each point is obtained by at least the average of 100 images. 
  \label{fig:LongitudeResults}}
\end{figure}

\begin{figure} 
\centerline{\includegraphics[width=0.7\textwidth]{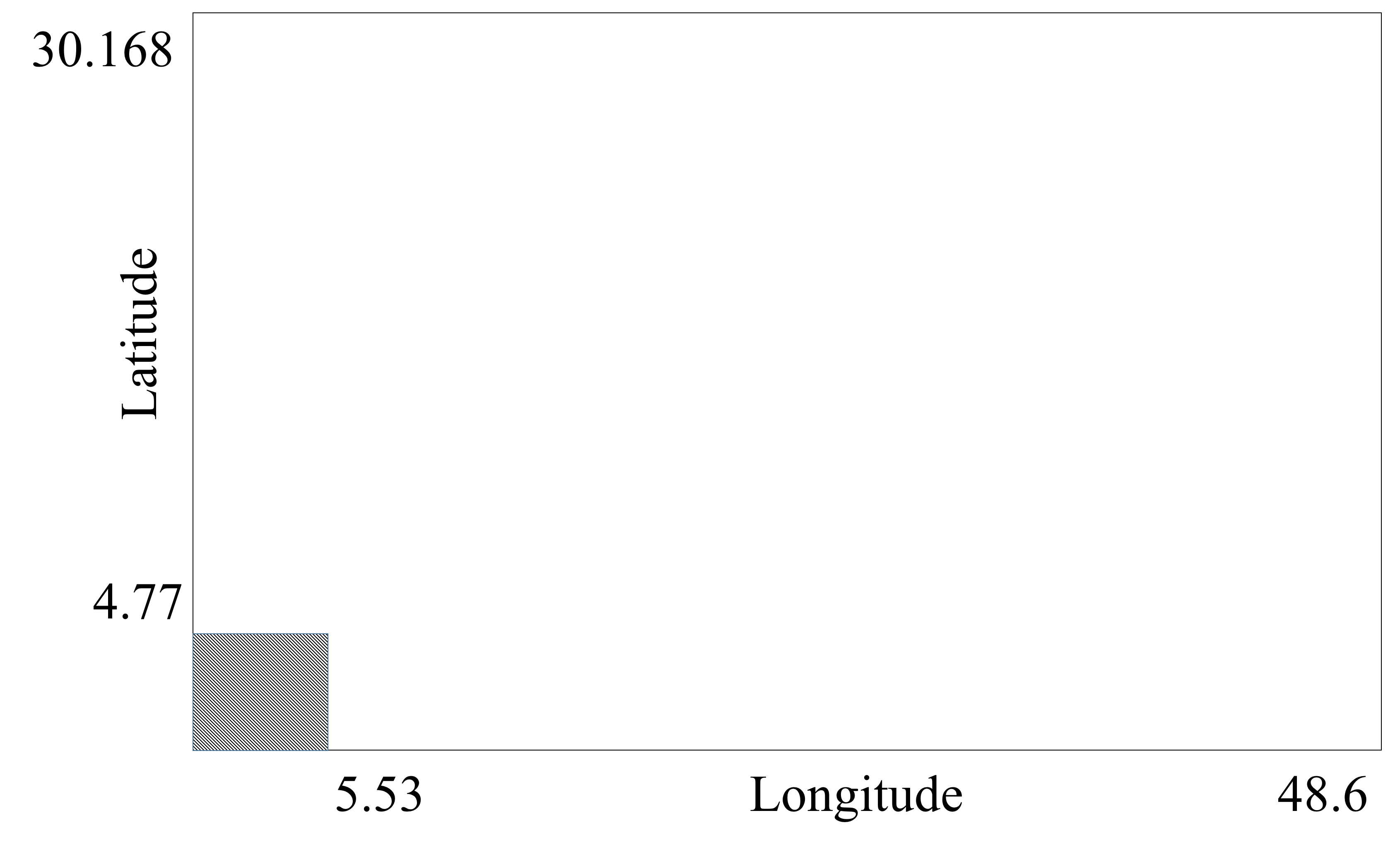}}
  \caption{Scale of the old and new error. For latitude error has been reduced from $30.168$ arcseconds to $4.77$ arcseconds. For longitude error has been reduced from $48.6$ arcseconds to $5.53$ arcseconds. 
  \label{fig:Reduced_error_shematic}}
\end{figure}

 In Figures~\ref{fig:old_latitude_error} and ~\ref{fig:old_longitude_error}, results are shown for the positioning without calculating $A_3$ are shown. In Figures~\ref{fig:LatitudeResults}  and ~\ref{fig:LongitudeResults}, results are shown for the positioning with calculating $A_3$. Average absolute deviation reduces for latitude and longitude, from $30.168$ to $4.77$ arcseconds and from $48.6$ to $5.53$ arcseconds, respectively ( Figure~\ref{fig:Reduced_error_shematic}).


\section{conclusion}
To get better results than about 5.5 arcseconds in longitude and 4.5 arcseconds in latitude (or approximately 200 meters), it is needed to be use a more accurate inclinometer. Instead on SCA-100T1, A701-2 from Jewell Instruments can be used which reduced the error 12.5 times but increases the price up to 20 times. On the other hand, the exposure time of the pictures taken from the sky needs to be decreased. Therefore it is needed a more advanced optics instruments with increased light gathering, which increases the price and of course makes harder the portability of the instrument. Therefore with the accessible facilities, this accuracy about 5.5 and 4.5 arcseconds are the most optimized results. This obtained accuracy is only a few times weaker than GPS \cite{kaartinen2015accuracy}, which is a very good alternative when GPS is not accessible or doesn't work well, which is great.
 
\bibliographystyle{spmpsci}
\bibliography{Local_Positioning_System_Using_Star_Tracker}

\begin{thebibliography}{10}
\providecommand{\url}[1]{{#1}}
\providecommand{\urlprefix}{URL }
\expandafter\ifx\csname urlstyle\endcsname\relax
  \providecommand{\doi}[1]{DOI~\discretionary{}{}{}#1}\else
  \providecommand{\doi}{DOI~\discretionary{}{}{}\begingroup
  \urlstyle{rm}\Url}\fi

\bibitem{hulbert2001accuracy}
Hulbert, I.A., French, J.: The accuracy of gps for wildlife telemetry and
  habitat mapping.
\newblock Journal of Applied Ecology \textbf{38}(4), 869--878 (2001)

\bibitem{wing2005consumer}
Wing, M.G., Eklund, A., Kellogg, L.D.: Consumer-grade global positioning system
  (gps) accuracy and reliability.
\newblock Journal of forestry \textbf{103}(4), 169--173 (2005)

\bibitem{pappalardi2001alternatives}
Pappalardi, F., Dunham, S., LeBlang, M., Jones, T., Bangert, J., Kaplan, G.:
  Alternatives to gps.
\newblock In: OCEANS, 2001. MTS/IEEE Conference and Exhibition, vol.~3, pp.
  1452--1459. IEEE (2001)

\bibitem{secroun2001high}
Secroun, A., Lampton, M., Levi, M.: A high-accuracy, small field of view star
  guider with application to snap.
\newblock Experimental Astronomy \textbf{12}(2), 69--85 (2001)

\bibitem{IzadmehrArXiv180600607I}
{Izadmehr}, M., {Khakian Ghom}, M.: {Design and construction of a high
  resolution, portable and low-cost positioner by a star tracking system}.
\newblock ArXiv e-prints  (2018)

\bibitem{wang2014}
Wang, B., Tian, L., Wang, Z., Yin, Z., Liu, W., Qiao, Q., Wang, H., Han, Y.:
  Image and data processing of digital zenith telescope (dzt-1) of china.
\newblock Chinese Science Bulletin \textbf{59}(17), 1984--1991 (2014)

\bibitem{wertz2012}
Wertz, J.R.: Spacecraft attitude determination and control, vol.~73.
\newblock Springer Science \& Business Media (2012)

\bibitem{shuster2004}
Shuster, M.D.: Deterministic three-axis attitude determination.
\newblock Journal of Astronautical Sciences \textbf{52}(3), 405--419 (2004)

\bibitem{markley1993attitude}
Markley, F.L.: Attitude determination using vector observations: A fast optimal
  matrix algorithm.
\newblock Journal of the Astronautical Sciences \textbf{41}(2), 261--280 (1993)

\bibitem{wahba1965least}
Wahba, G.: A least squares estimate of satellite attitude.
\newblock SIAM review \textbf{7}(3), 409--409 (1965)

\bibitem{bochkanov2011alglib}
Bochkanov, S., Bystritsky, V.: Alglib-a cross-platform numerical analysis and
  data processing library.
\newblock ALGLIB Project. Novgorod, Russia  (2011)

\bibitem{goodall199313}
Goodall, C.R.: 13 computation using the qr decomposition  (1993)

\bibitem{kaartinen2015accuracy}
Kaartinen, H., Hyypp{\"a}, J., Vastaranta, M., Kukko, A., Jaakkola, A., Yu, X.,
  Py{\"o}r{\"a}l{\"a}, J., Liang, X., Liu, J., Wang, Y., et~al.: Accuracy of
  kinematic positioning using global satellite navigation systems under forest
  canopies.
\newblock Forests \textbf{6}(9), 3218--3236 (2015)

\end{thebibliography}

\end{document}